\documentclass[AMA,STIX1COL]{WileyNJD-v2}

\usepackage{subcaption}

\articletype{Article Type}%

\received{}
\revised{}
\accepted{}
        
\begin{document}

\title{Computing Viscous Flow Along a 3D Open Tube Using the Immerse Interface Method}

\author[1]{Sarah E. Patterson*}
\author[2]{Anita T. Layton}

\authormark{PATTERSON \textsc{et al}}

\address[1]{\orgdiv{Applied Mathematics Department}, \orgname{Virginia Military Institute}, \orgaddress{\state{Virginia}, \country{United States}}}
\address[2]{\orgdiv{Departments of Applied Mathematics and Biology, Cheriton School of Computer Science, and School of Pharmacy}, \orgname{University of Waterloo}, \orgaddress{\state{Ontario}, \country{Canada}}}

\corres{*Sarah Patterson, \orgdiv{Applied Mathematics Department}, \orgname{Virginia Military Institute}, \orgaddress{\state{Virginia}, \country{United States}}. \email{pattersonse@vmi.edu}}


\abstract[Summary]{In a companion study \cite{patterson2020computing2D}, we present a numerical method for simulating 2D viscous flow through an open compliant closed channel, drive by pressure gradient. We consider the highly viscous regime, where fluid dynamics is described by the Stokes equations, and the less viscous regime described by the Navier-Stokes equations. In this study, we extend the method to 3D tubular flow. The problem is formulated in axisymmetric cylindrical coordinates, an approach that is natural for tubular flow simulations and that substantially reduces computational cost. When the elastic tubular walls are stretched or compressed, they exert forces on the fluid. These singular forces introduce unsmoothness into the fluid solution.
As in the companion 2D study \cite{patterson2020computing2D}, we extend the immersed interface method to an open tube, and we compute solution to the model equations using the resulting method. Numerical results indicate that this new method preserves sharp jumps in the solution and its derivatives, and
converges with second-order accuracy in both space and time.}


\maketitle


\section{Introduction}

A detailed description of 3D viscous fluid flow through a compliant or actively moving tube is of interest in many biological applications including pumping via peristalsis in a valveless heart \cite{santhanakrishnan2011fluid}, food mixing in the intestine \cite{tharakan2010mass}, and blood flow through a vessel \cite{tang2002simulating}.  Computer simulations that assume rigid walls often fail to predict some essential characteristics of tubular flow, such as pressure wave propagation. Therefore, such simulations cannot be considered reliable in every situation, such as the case when the vessels undergo relatively large displacements. 

 A natural way to model flow in compliant vessels is to frame it as an immersed boundary problem \cite{arthurs1998modeling, peskin2002immersed}.
 Immersed boundary problems are a subset of FSI problems in which a thin structure or physical boundary is present in the fluid \cite{hou2012review}.  The immersed boundary formulation was developed by Charles Peskin to study blood flow in the heart but has been extended to a variety of biological applications including blood clotting, aquatic animal locomotion, fluid dynamics in the inner ear, arteriolar flow, and flow in collapsible tubes \cite{arthurs1998modeling, peskin1995general}. The popularity of this method is due to its ability to model fluid interactions with complex, passive, or active elastic material with relative ease compared to traditional body fitted approaches.  The fluid solution is computed on a fixed Cartesian grid which does not need to conform to the geometry of the structure. The structure can then be represented using Lagrangian variables that can move over the fluid grid unimpeded. The Eulerian and Lagrangian variables are related by interaction equations that contain the delta Dirac function \cite{mittal2005immersed, peskin2002immersed}.

 The immersed boundary method generally only converges with first-order accuracy and does not capture the discontinuities in the velocity gradient and pressure near the interface.  Since the vessel movement is determined from the local fluid velocity, these inaccuracies cause the movement of the interface to deteriorate over time and lose volume. Although progress has been made on improving the volume conservation of the immersed boundary method, this problem has not been eliminated \cite{peskin1993improved}.  Additionally, the pressure and velocity near the interface are needed for some applications. A sharp interface method, like the immersed interface method, is required in order to determine these variables accurately.
 
 The immersed interface method overcomes the drawbacks in immersed boundary methods by sharply capturing the discontinuities of fluid solutions and the movement of the immersed structures. It is based on a method used to compute solutions to Poisson's equation in irregular domains developed by Anita Mayo \cite{mayo1984fast}. The immersed interface method has been applied to Poisson's equation \cite{leveque1994immersed}, the Stokes equations \cite{leveque1997immersed}, and the Navier-Stokes equations \cite{li2001immersed,lee2003immersed}. The immersed interface method achieves second-order accuracy in numerical approximation. 
For fluid-structure interactions problems, the immersed interface method has been used to compute the coupled motion of a viscous fluid and a simple closed elastic interface with second-order accuracy  \cite{le2006immersed,lee2003immersed,li2014computing,li2001immersed,tan2009immersed,xu2006immersed}.

 In the original derivation of the jump conditions, the interface was a closed surface. Therefore,  blood vessels have been modeled as a closed interface in the shape of a tube with capped ends \cite{arthurs1998modeling,li2014computing,li2013hybrid,rosar2001fluid,smith2003advective,rosar1994three}. The closed tube was immersed inside a rectangular fluid domain.  The flow in the closed tube is driven by adding a fluid source and sink to opposing ends of the tube. These modifications create unrealistic flow in biological models of blood vessels, especially near the source and sink. Typically, a large computational domain is used to compute solutions, but only the results in the center of the tube away from the source and sink are considered.   We have created a novel extension of the immersed interface method to interfaces that are not closed, but instead, are shaped like an open tube that spans from one end of the fluid domain to the other. This method will be referred to as the immersed interface method for open tubes (IIM-OT). By using the IIM-OT, we can create a more natural fluid profile using a smaller computational domain.

Section 2 details the problem formation. In particular, the immersed interface method requires the magnitudes of the discontinuities or jump conditions of the primitive variables, and their derivatives are determined a priori. We show that the jump conditions for the immersed interface method for both closed and open tube-shaped interface can be computed in the same manner from the force strength function. 
Section 3 describes the numerical method. 
Section 4 includes numerical simulations to show that the novel IIM-OT for the Navier-Stokes equations and the Stokes equations achieved second-order accuracy in space for both 2D simulations in rectangular coordinates and 3D simulations in axisymmetric cylindrical coordinates.

\section{Problem formulation}
\label{sec:StokesFormulation}
\subsection{Computational domain and immersed interface}
We formulate a model that simulates fluid flow through an open tube with compliant walls, extending from one boundary of the computational domain to the opposite boundary. 
We formulate the 3D model in cylindrical coordinates in the fluid domain $\Psi=[0,H]\times[0,2\pi]\times[0,L]$ which contains an immersed interface $\Gamma$ in the shape of an infinitely-thin, compliant tube with a nonuniform diameter that extends from one end of the domain to the opposing end, shown in Fig. \ref{fig:cyl3Ddomain}. Additionally, we assume that the flow is axisymmetric, thereby allowing us to reduce the computational costs from 3D to 2D. The 2D computational domain is a slice of the 3D domain where $\theta=0$ or $\Psi|_{\theta=0}=[0,H]\times[0,L]$ , shown in Fig. \ref{fig:cyl2Ddomain}. The immersed interface surface $\Gamma$ can be represented as a 2D curve, which is also be called $\Gamma$  since the structure of interest is clear from the context. Assume $\Gamma$ intersects the computational domain boundary at $(r,z)=(a,0)$ and  $(r,z)=(b,L)$. Let $\mathbf{n}$ be the unit normal vector oriented towards the outside of the tube.  \\

\begin{figure}[ht!]
\centering
\begin{tabular}{cc}
 \includegraphics[]{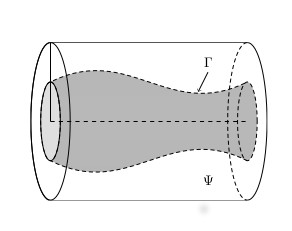} &
 \includegraphics[]{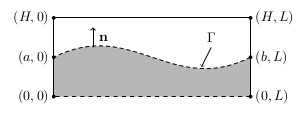} \vspace{-.2in} \\
 a & b
\end{tabular}
\caption[The open immersed interface, $\Gamma$, in the 3D axisymmetric fluid domain and the 2D computational domain.]{(a) The 3D  fluid domain $\Psi=[0,H]\times[0,2\pi]\times[0,L]$ is radially symmetric and represented in cylindrical coordinates $(r,\theta,z)$. The open interface $\Gamma$ is a surface in the shape of an irregular tube that spans $\Psi$. (b) The 2D computational domain is a slice of the 3D domain where $\theta=0$ or $\Psi|_{\theta=0}=[0,H]\times[0,L]$. The computational slice contains an open immersed interface curve $\Gamma$. Note $\Gamma$ intersects the computational domain boundary at $(r,z)=(a,0)$ and  $(r,z)=(b,L)$.  $\mathbf{n}$ is the unit normal oriented towards the outside of the tube. }
\label{fig:cylDomain}
\end{figure}

\subsection{Boundary condition} 

Appropriate boundary conditions must be chosen to drive the fluid through the tube, which can be done by creating an axial pressure gradient inside the vessel. One way to accomplish this is to specify the value of pressure on the inlet and outlet of the tube using Dirichlet boundary conditions. 
 
At the inlet, it is assumed that $w$ has a parabolic profile consistent with 3D Poiseuille flow.  In 3D axisymmetric cylindrical coordinates, the pressure gradient and velocity are related by
\begin{equation}w=-\frac{dp}{dz}\frac{1}{4\mu}(R^2-y^2)\end{equation}
\begin{equation}u=-\frac{dp}{dr}=0.\end{equation}  
In the velocity decomposition method, boundary conditions must be imposed for both the Stokes and regular parts such that the boundary conditions for the full solution are still satisfied. The boundary conditions for the full solution are non-smooth for $w$ and discontinuous for $p$ at the inlet of the tube. The boundary conditions for the full solution are imposed on the Stokes part since $w_s$ is non-smooth and $p_s$ is discontinuous near the boundary.  Homogeneous boundary conditions of the same type as the full solution are imposed for the regular solution. Below are the boundary conditions used for the two simulations in Section \ref{sec:NSCylResults} where the average inlet velocity is denoted $\bar{w}$. 
\begin{itemize}
\item Inlet Boundary ($z=0$)
\begin{equation}
    p|_{z=0}=p_s|_{z=0}=\begin{cases} 
      0 & r> R \\
     \frac{ 8\mu\bar{w}L}{R^2} &  r\leq R 
   \end{cases},
\end{equation}
\begin{equation}
       w|_{z=0}=w_s|_{z=0}=\begin{cases} 
      0 & r>R \\
     \frac{ 2\bar{w}(R^2-r^2)}{R^2} & r \leq R 
   \end{cases},
\end{equation}
\begin{equation}
   u|_{z=0}=u_s|_{z=0}=0
   \end{equation}
\begin{equation}
   p_r|_{z=0}=u_r|_{z=0}=w_r|_{z=0}=0
   \end{equation}
\item Outlet Boundary ($z=L$)
\begin{equation}
p|_{z=L}=p_s|_{z=L}=p_r|_{z=L}=0 \end{equation}
\begin{equation}
 \frac{\partial w}{\partial z}\Big|_{z=L}=\frac{\partial w_s}{\partial z}\Big|_{z=L}=\frac{\partial w_r}{\partial z}\Big|_{z=L}=0
\end{equation}
\begin{equation}
    \frac{\partial u}{\partial z}\Big|_{z=L}=\frac{\partial u_s}{\partial z}\Big|_{z=L}=\frac{\partial u_r}{\partial z}\Big|_{z=L}=0
\end{equation}

\item Top Boundary ($r=H$)
\begin{equation}
\frac{\partial p}{\partial r}\Big|_{r=H}=\frac{\partial p_s}{\partial r}\Big|_{r=H}=\frac{\partial p_r}{\partial r}\Big|_{r=H}=0 \end{equation}
\begin{equation}
 \frac{\partial w}{\partial r}\Big|_{r=H}=\frac{\partial w_s}{\partial r}\Big|_{r=H}=\frac{\partial w_r}{\partial r}\Big|_{r=H}=0
\end{equation}
\begin{equation}
    \frac{\partial u}{\partial r}\Big|_{r=H}=\frac{\partial u_s}{\partial r}\Big|_{r=H}=\frac{\partial u_r}{\partial r}\Big|_{r=H}=0
\end{equation}
\item Bottom Boundary ($r=0$)
\begin{equation}
\frac{\partial p}{\partial r}\Big|_{r=0}=\frac{\partial p_s}{\partial r}\Big|_{r=0}=\frac{\partial p_r}{\partial r}\Big|_{r=0}=0 \end{equation}
\begin{equation}
 \frac{\partial w}{\partial r}\Big|_{r=0}=\frac{\partial w_s}{\partial r}\Big|_{r=0}=\frac{\partial w_r}{\partial r}\Big|_{r=0}=0
\end{equation}
\begin{equation}
u|_{r=0}=u_s|_{r=0}=u_r|_{r=0}=0
\end{equation}
\end{itemize}

\subsection{Fluid structure interactions}

The Navier-Stokes equations take the form 
\begin{eqnarray}
\rho\left(\frac{\partial \mathbf{u}}{\partial t}+\mathbf{u}\cdot\nabla\mathbf{u}\right)&=&\mu\Delta\mathbf{u}-\nabla p +\mathbf{F}\label{eq:NS1b}\\
\nabla \cdot {\mathbf u}&=&0 \label{eq:NS2b}
\end{eqnarray}
\noindent
where  ${\mathbf u}=(u,v,w)$, where $u$, $v$, and $w$ denote velocity components in the $r$-, $\theta$-, and $z$-directions, respectively, $p$ is the pressure, and $\mu$ is viscosity; and $\mathbf{F}$ is the interfacial force, which is singularly supported along $\Gamma$ (see below).  We connsider also the zero Reynolds number regimes given by the Stokes equations:
\begin{equation}
\mu\Delta\mathbf{u}-\nabla p +\mathbf{F} = 0\label{eq:Stokes}
\end{equation}
and the continuity equation (Eq.~\ref{eq:NS2b})

The force exerted by the interface $\Gamma$ can be written as  
\begin{equation}
{\bf F}({\bf x},t)=\int\displaylimits_{\Gamma}{\bf f}(\alpha)\delta({\bf x} - {\bf X}(\alpha))d\alpha,
\label{eq:F}
\end{equation}
where $\mathbf{X}$ denotes the position of the interface,  $\alpha=\alpha(s,\theta,t)$ in 3D is the material
coordinate(s) that parameterize the interface curve or surface at time $t$, ${\bf f}(\alpha)$ is the force strength at the point ${\bf X}(\alpha)$, and $\delta$ is the Dirac delta function.   Since all 3D cases are axisymmetric, the interface is effectively a 2D curve, and it is sufficient to only consider $\alpha=\alpha(s,t)$.


The force strength from Eq. \ref{eq:F}
\begin{equation}
{\bf f}(\alpha)={\mathbf f}_{E}(\alpha) + {\mathbf f}_{T}(\alpha).
\end{equation}
is comprised of a two major components: an elastic force ${\mathbf f}_{E}$ and a tether force ${\mathbf f}_{T}$.  Since the interface has elastic properties, deviation from its resting configuration generates a restorative force. 
In 3D cylindrical coordinates, the elastic force derived in Ref.~ \cite{lai2011simulating} is given by
\begin{equation}
{\bf f}_E(s,t)=\frac{\partial T}{\partial \mathbf{\tau}}\mathbf{\tau}-2T\kappa\mathbf{n}
\label{eq:Felastic1}
\end{equation}
where $\kappa$ is the mean curvature and the tension $T(s,t)$ is given by
\begin{equation}
T(s,t)=a_E\left( \left|\frac{\partial {\bf X}}{\partial s}\right|-1\right)
\label{eq:Felastic2}
\end{equation}
In Eq.~\ref{eq:Felastic2}, $a_E$ controls the stiffness of the interface.
In Eq.~\ref{eq:Felastic1}, the unit tangent vector to $\Gamma$ is given by  
\begin{equation}
\tau (s,t)=\frac{\partial {\bf X}/\partial s}{\left|\partial {\bf X}/\partial s\right|}.
\label{eq:Felastic3}
\end{equation}

We assume that at steady state with the tethers at their equilibrium positions, tubular flow is that of Poiseuille flow, with fluid outside of the channel at rest. Thus, at steady state, one expects a jump discontinuity in p and in $u_n$ across $\Gamma$. To generate this steady-state flow profile, we assignn to the tether forces  two components:
\begin{equation}
{\mathbf f}_T = {\mathbf f}_T^0 + {\mathbf f}'_T 
\end{equation}
where ${\mathbf f}_T^0$ is the force component needed to support Poiseuille flow along the channel (by generating the necessary jumps in the solution and its derivative). 

The second component ${\mathbf f}_T'$ arises from the displacement of the tethers from their equilibrium positions or anchor points.
The interface control knots are tethered to anchor points in the fluid domain by a spring with resting length 0.  Suppose the interface ${\mathbf X}$ is anchored to $\bar{{\mathbf X}}$.
Let $a_{T}$ be the spring force constant. 
Then tether force is 
\begin{equation}
{\mathbf f}_{T}'=a_{T}(\bar{{\mathbf X}}-{\mathbf X}). 
\end{equation}
If the interface knots move away from their anchor points, restorative forces are generated. Interface movement can either be restricted by using stationary anchor points or induced by moving the position of the anchor points in time.

The interface is deformable and is assumed to move at the same speed as the local fluid. The no-slip condition
\begin{equation}
\frac{d\mathbf{X}}{dt}=u(\mathbf{X}).
\label{eq:NoSlip}
\end{equation}
\noindent
 describes this motion.

\subsection{Boundary conditions}
Appropriate boundary conditions must be chosen to drive the fluid through the tube, which can be done by creating an axial pressure gradient in the vessel. One way to accomplish this is to specify the value of pressure on the inlet and outlet of the tube using Dirichlet boundary conditions. Another way is to impose inhomogeneous periodic boundary conditions.
Specific boundary conditions used for the 2D and 3D cases simulations are discussed before the results.

Bi-periodic boundary conditions are imposed for velocity, which implies that the volume of the fluid in the tube remains constant in time.  To drive flow, we prescribe a pressure gradient inside the tube, by requiring there to be a constant difference in pressure $P_{diff}$ at the inlet and outlet of the tube
 \begin{equation}
   p(0,y)=p(L,y)-  \begin{cases}
   0 & |y|\geq R\\
P_{diff} & |y|<R 
     \end{cases} 
     \label{eq:inhomoPeriodicBC1}
     \end{equation}
     \noindent
 where $R$ is the radius of the tube.   The derivatives of pressure on the $x=0$ and $x=L$ boundaries are required to be equal
 \begin{equation}
  \frac{\partial p}{\partial x}|_{x=0}= \frac{\partial p}{\partial x}|_{x=L}. 
  \label{eq:inhomoPeriodicBC2}
 \end{equation}
 \noindent
 This is referred to as an inhomogeneous periodic boundary condition. 
Not only does this create a pressure gradient in the tube, but it also forces the pressure gradient to be periodic, which is needed since periodic boundary conditions are imposed for velocity. Periodic boundary conditions are imposed for pressure at the $y=\pm H$ boundaries.

\subsection{Derivation of jump conditions}\label{subsec:derivationJump}

The jump conditions for the Stokes equations and the Navier-Stokes equations are derived in Ref.~ \cite{li2013hybrid} and Ref.~\cite{li2014computing} respectively for closed interfaces. Both the Stokes equations and the Navier-Stokes equations have the same jump conditions given the same interfacial forces \cite{li2013hybrid,li2014computing}.  By creating a fictitious closed interface by adding segments to close the open tube interface, the same derivation for closed interfaces from Ref.~\cite{li2014computing} can be applied to open tube-shaped interfaces. This is shown in detail in Section \ref{subsec:derivationJump} for the Stokes equations. The jump conditions for both the 3D Stokes and Navier-Stokes equations in cylindrical coordinates for both closed and open tube-shaped interfaces are 
\begin{align}
[p]&= f_n\\
\left[\frac{\partial p}{\partial n}\right]&=\frac{1}{r}\frac{\partial(r f_s)}{\partial s}\\
\left[\mu\frac{\partial u}{\partial n}\right]&=f_s \sin(\alpha)\\
\left[\mu\frac{\partial w}{\partial n}\right]&=-f_s \cos(\alpha)
\end{align}
where $\alpha$ is the angle between the normal and $r$ direction  \cite{li2013hybrid,li2014computing}.

Next, we derive the jump conditions across an open tube for the 3D Stokes equations in axisymmetric cylindrical coordinates. We show that the jump conditions in cylindrical coordinates for an interface that is in the shape of an open tube have the same dependence on force and interface position as in the closed tube case.

 Again, the interface must be shaped like a closed surface for the derivation. Therefore, consider the fictitious closed surface 
\begin{equation}\Gamma_c=\Gamma\cup \{(r,\theta,0) | r\in[0,a], \theta\in[0, 2\pi]\}\cup \{(r,\theta,L) | r \in[0,b], \theta\in[0,2\pi]\}\end{equation} 
which is formed by the interface $\Gamma$ and the inlet and outlet of the tube which intersect the fluid domain $\Psi$ at $z=0$ and $z=L$ respectively.

Let $\Omega_{\epsilon}$ be an extended domain that contains the surface interface $\Gamma_c$ where the distance between $\Omega_{\epsilon}$ and $\Gamma_c$ shrinks to zero as $\epsilon$ approaches zero. Note that $\Omega_{\epsilon}$ extends outside of the original fluid domain $\Psi$.

 Although symmetry is imposed later, first consider the fully 3D Stokes equations in cylindrical coordinates, given by 
\begin{equation} \frac{\partial p}{\partial r}=\mu\left( \Delta-\frac{1}{r^2}\right)u-\frac{2\mu}{r^2}\frac{\partial v}{\partial\theta}+\hat{F_1}
\label{eq:FullStokesCyl1}
\end{equation}
\begin{equation} \frac{1}{r}\frac{\partial p}{\partial \theta}=\mu\left(   \Delta-\frac{1}{r^2}\right)v+\frac{2\mu}{r^2}\frac{\partial u}{\partial\theta}+\hat{F_2}
\label{eq:FullStokesCyl2}
\end{equation}
\begin{equation}
\frac{\partial p}{\partial z}=\mu \Delta w+\hat{F_3}
\label{eq:FullStokesCyl3}
\end{equation}
\begin{equation}
\frac{1}{r}\frac{\partial (r u)}{\partial r}+\frac{1}{r}\frac{\partial v}{\partial\theta}+\frac{\partial w}{\partial z}=0
\label{eq:FullStokesCyl4}
\end{equation}
where $\mathbf{x}=(r,\theta,z)$, $u$, $v$, and $w$ are the velocities in the $r$, $\theta$, and $z$ directions respectively, and $\mathbf{\hat{F}}=(\hat{F_1},\hat{F_2},\hat{F_3})$ with
\begin{equation}
\hat{F_i}=\iint_{\Gamma_c}\hat{f}_i(s,\theta,t)\delta(\mathbf{x}-\mathbf{X}(s,\theta,t))dS     
\end{equation} for $i=1,2,3$, where $\mathbf{\hat{f}}$ is the extension of $\mathbf{f}$ which is supported on $\Gamma$ and agrees with $\mathbf{f}$ on $\Gamma$. Even if the correction terms are computed for the fictitious closed domain $\Gamma_c$ there is no contribution from the extensions of the interface. Therefore, the hat symbol is dropped for the rest of the derivation. Now the derivation mirrors the procedure found in Ref. \cite{li2013hybrid}. 

Taking the divergence of Eqs. \ref{eq:FullStokesCyl1}, \ref{eq:FullStokesCyl2}, and \ref{eq:FullStokesCyl3} gives Poisson's equation for pressure
\begin{equation}
    \Delta p =\nabla\cdot \mathbf{F}.
    \label{eq:FullStokesCyl5}
\end{equation}
Let $\phi(r,\theta,z)$ be an arbitrary twice continuously differentiable test function. Multiply the left hand side of Eq. \ref{eq:FullStokesCyl5} by $\phi$ and integrate over $\Omega_{\epsilon}$
\begin{align}
    \iiint_{\Omega_{\epsilon}} \Delta p\phi dV 
    &=\iint_{\partial \Omega_{\epsilon}} \phi(\nabla p \cdot\mathbf{n}) dS -\iiint_{\Omega_{\epsilon}} \nabla\phi\cdot\nabla p dV \\
  &=\iint_{\partial \Omega_{\epsilon}} \phi(\nabla p \cdot\mathbf{n}) dS -\left(\iint_{\partial\Omega_{\epsilon}}(\nabla\phi\cdot\mathbf{n})p dS -\iiint_{\Omega_{\epsilon}} (\Delta\phi)p dV\right) \\
    &\to \iint_{\Gamma_c} \phi\left[\frac{\partial p}{\partial \mathbf{n}} \right]dS -\iint_{\Gamma_c}\frac{\partial \phi}{\partial \mathbf{n}}[p] dS +0 \textrm{   as }\epsilon\to 0. 
\end{align}
The first and second lines are by Green's first Identity and the divergence theorem, respectively. The last line is found by taking the limit as $\epsilon$ goes to zero. The last term goes to zero since pressure is bounded. The surface element is $dS=rdsd\theta$. By imposing the axisymmetric assumptions on $p$ and $\phi$
we get that 
\begin{equation}
   \iiint_{\Omega_{\epsilon}} \Delta p\phi dV \to 2\pi\left(\int_{\Gamma_c} \phi\left[\frac{\partial p}{\partial \mathbf{n}} \right]rds -\int_{\Gamma_c}\frac{\partial \phi}{\partial \mathbf{n}}[p] rds\right), 
\end{equation}
where $\Gamma_c$ denotes the curve formed by restricting the surface $\Gamma_c$ to a fixed $\theta$. 
Multiply the right hand side of Eq. \ref{eq:FullStokesCyl5} by $\phi$ and integrate over $\Omega_{\epsilon}$.
\begin{align}
    \iiint_{\Omega_{\epsilon}} (\nabla\cdot \mathbf{F})\phi(\mathbf{x}) dV 
    &=\iiint_{\Omega_{\epsilon}}\left( \iint_{\Gamma_c}\nabla\cdot \left(\mathbf{f}(s,\theta,t)\delta(\mathbf{x}-\mathbf{X}(s,\theta,t))\right)dS\right)\phi(\mathbf{x}) dV\\
    &= \iint_{\Gamma_c}\mathbf{f}(s,\theta,t)\left(\iiint_{\Omega_{\epsilon}}\nabla\cdot \delta(\mathbf{x}-\mathbf{X}(s,\theta,t))\phi(\mathbf{x}) dV\right)dS\\
    &= -\iint_{\Gamma_c} \mathbf{f}(s,\theta,t)\cdot\nabla\phi(\mathbf{x}) dS \label{eq:GradForceInt4}
\end{align}
The last line is found using the divergence theorem and noting that the $\delta(\mathbf{x}-\mathbf{X}(s,\theta,t))$ has support on $\Gamma_c$ and is zero on $\partial \Omega_{\epsilon}$.
Applying the axisymmetric assumption that $\phi$ and $f$ are independent of $\theta$ to Eq. \ref{eq:GradForceInt4} yields
\begin{align}
    \iiint_{\Omega_{\epsilon}} (\nabla\cdot \mathbf{F})\phi(\mathbf{x}) dV 
    &= -2\pi\int_{\Gamma_c}\left( f_1\frac{\partial\phi}{\partial r}+f_3\frac{\partial\phi}{\partial z}\right)rds\\
        &= -2\pi\int_{\Gamma_c}\left( f_n\frac{\partial\phi}{\partial n}+f_s\frac{\partial\phi}{\partial s}\right)rds\\
        &= -2\pi\int_{\Gamma_c}\left( f_nr\frac{\partial\phi}{\partial n}-\frac{\partial f_s}{\partial s}\phi\right)ds
    \end{align}
    where $\Gamma_c$ denotes the curve formed by restricting $\Gamma_c$ to a fixed $\theta$. The second line is found by breaking the force into components normal and tangential to the interface. The last line is found using integration by parts.
Since $\phi$ is arbitrary, we have
\begin{align}
    [p]&=f_n\\
    \left[\frac{\partial p}{\partial n} \right]&=\frac{1}{r}\frac{\partial (f_s r)}{\partial s}.
\end{align}

To find the jump conditions for $u$, multiply Eq. \ref{eq:FullStokesCyl1} by $\phi$ and integrate
\begin{align}
  \iiint_{\Omega_{\epsilon}}\phi\frac{\partial p}{\partial r}dV = \iiint_{\Omega_{\epsilon}}\phi\mu\Delta udV 
  - \iiint_{\Omega_{\epsilon}} \phi\frac{\mu u}{r^2}dV 
  + \iint_{\Gamma_c}f_1\phi dS.  \label{eq:cylJumpDeriveUint}
\end{align}

The first term on the right hand side of the Eq. \ref{eq:cylJumpDeriveUint} approaches the following as $\epsilon\to 0:$
\begin{align}
    \iiint_{\Omega_{\epsilon}}\phi\mu\Delta udV&= \iint_{\partial\Omega_{\epsilon}}\phi\mu(\nabla u\cdot \mathbf{n})dS-
    \iiint_{\Omega_{\epsilon}}\mu \nabla\phi\cdot\nabla u dV\\
    &\to \iint_{\Gamma_c}\phi\left[\mu\frac{\partial  u}{\partial \mathbf{n}}\right]dS-0
\end{align}

The second term on the right hand side of the Eq. \ref{eq:cylJumpDeriveUint} 
\begin{align}
    \iiint_{\Omega_{\epsilon}} \phi\frac{\mu u}{r^2}dV =  \iiint_{\Omega_{\epsilon}} \phi\frac{\mu u}{r^2}rdrd\theta dz= \iiint_{\Omega_{\epsilon}} \phi\frac{\mu u}{r}drd\theta dz 
\end{align}
approaches 0 as $\epsilon\to 0$. This is true because by the axisymmetric condition when $r\to 0$, $u$ approaches zero. Additionally $\lim_{r\to0} \frac{u}{r}=\frac{\partial u}{\partial r}$, which is finite. 

The first term on the left hand side of the Eq. \ref{eq:cylJumpDeriveUint} approaches the following as $\epsilon\to 0$.
\begin{align}
  \iiint_{\Omega_{\epsilon}}\phi\frac{\partial p}{\partial r}dV &=   \iiint_{\Omega_{\epsilon}}\phi\nabla\cdot[ p,0]-\frac{p}{r}dV\\
  &\to \iint_{\Gamma_c}\phi[p]\cos(\alpha)dS + 0
\end{align}
where $\alpha$ is the angle between the normal and $r$ direction. 
Therefore 
\begin{equation}
\left[\mu\frac{\partial u}{\partial n}\right]=[p]\cos(\alpha)-\frac{f_1}{|f_1|}=f_s\sin(\alpha). 
\end{equation}
Similarly, 
\begin{equation}
\left[\mu\frac{\partial w}{\partial n}\right]=[p]\sin(\alpha)-\frac{f_3}{|f_3|}=f_s\cos(\alpha). 
\end{equation}


\section{Numerical method}
\label{sec:StokesNumerics}

To solve the immersed boundary problem, we compute the fluid velocity and pressure on a fixed Eulerian grid. A
 moving Lagrangian frame of reference is used to track the location of the interface $\Gamma$ over time. In 2D, the Stokes equations are solved on the fluid computational grid, 
\[\Omega_h=\left\{\mathbf{x}_{i,j}=\left(jh,ih-\frac{H}{2}\right) | i\in{1,\dots, N_y} \textrm{ and } j\in{1,\dots, N_x}\right\}\]
where $h$ is the grid spacing and $N_x=\frac{L}{h}+1$ and $N_y=\frac{H}{h}+1$ are the number of grid points in the $x$ and $y$ directions respectively. In 3D, the Stokes equations are solved on the fluid computational grid, 
\[\Omega_h=\left\{\mathbf{x}_{i,j}=\left(ih,jh-\frac{H}{2}\right) | i\in{1,\dots, N_r} \textrm{ and } j\in{1,\dots, N_z}\right\}\]
where $h$ is the grid spacing and $N_r=\frac{H}{h}+1$ and $N_z=\frac{L}{h}+1$  are the number of grid points in the $r$ and $z$ directions respectively. In both the 2D and 3D case, the interface position at time $t_n=n\Delta t$ is tracked by $N_b$ boundary markers $\mathbf{X}^{n}=\{\mathbf{X}_i^{n}\}_{i=1}^{N_b}$ that are connected by a cubic spline.

\subsection{Velocity decomposition method}
\label{sec:VDecomp}

The velocity decomposition method, developed by Ref.~ \cite{beale2009velocity}, can be used to avoid computing most of the correction terms that are needed for the immersed interface method for the Navier-Stokes equations.
This method leverage the fact that, given the same singular interfacial force, the jumps in the solutions are identical for both the Stokes equations and the Navier-Stokes equations as shown by Ref.~ \cite{leveque1997immersed} and Ref.~ \cite{li2001immersed} respectively. This result occurs because the no-slip condition implies that velocity and its material derivative are continuous at the interface \cite{lai2001remark}. 

It is sufficient to break the solution of the Navier-Stokes equations into a part that satisfies the Stokes equations with the singular interfacial forces and a regular part that solves the remaining equations as shown below.

Suppose $\mathbf{u}$ and $p$ solve the Navier-Stokes equations with a singularly-supported interfacial force $\mathbf{F}$. Let
\begin{equation} \mathbf{u}=\mathbf{u}_s+\mathbf{u}_r\end{equation}
\begin{equation}{p}={p_s}+{p_r}\end{equation}
\noindent
where the subscripts $s$ and $r$ are used for the Stokes and regular part, respectively. 
The Stokes part, $\mathbf{u}_s$ and $p_s$, satisfies the Stokes equations with the interfacial force $\mathbf{F}$
\begin{equation} 
-\mu\Delta\mathbf{u_s}=-\nabla p_s +\mathbf{F} \label{eq:stokepart1}
\end{equation}
\begin{equation}\nabla\cdot \mathbf{u}_s=0.\end{equation}
\noindent
The regular part, $\mathbf{u}_r$ and $p_r$, satisfies the remaining terms in the full Navier-Stokes equations
\begin{equation} 
\rho\left(\frac{\partial \mathbf{u}_r}{\partial t}+\mathbf{u}\cdot\nabla\mathbf{u}_r\right)=\mu\Delta\mathbf{u_r}-\nabla p_r +\mathbf{F_b}
\label{eq:NSregular1}\end{equation}
\begin{equation}
\nabla\cdot \mathbf{u}_r=0
\label{eq:NSregular2}\end{equation}
\noindent
where 
\begin{equation}
\mathbf{F}_b=-\rho\left(\frac{\partial \mathbf{u}_s}{\partial t}+\mathbf{u}\cdot\nabla\mathbf{u}_s\right).
\label{eq:NSregular3}\end{equation}
\noindent
The Stokes part is found first, so $\mathbf{F}_b$ is known and treated as a body force when solving for the regular part. The force $\mathbf{F}_b$ is a continuous function in the fluid domain since $\mathbf{u}_s$ and its material derivative are continuous. Notice that the full velocity $\mathbf{u}$ is used in the transport of $\mathbf{u}_r$ and in $\mathbf{F}_b$.

This method has several advantages over the immersed interface method for the Navier-Stokes equations. First, correction terms only need to be computed for the Stokes part of the solution, saving considerable time in implementation since numerous correction terms are needed for the additional terms in the Navier-Stokes equations. Secondly, the velocity decomposition method can also be implemented with fractional time steps that help overcome some of the challenges of simulating stiff interfaces. Small fractional time steps are used to move the interface, which is done by only computing the velocity on the interface. This computation is done using Stokeslets to compute the Stokes part of the velocity at the interface and interpolating the regular part of velocity on the interface in time \cite{beale2009velocity}. Therefore, the interface position can be updated without computing the full solution at each fractional time step.  

\subsection{Stokes part: immersed interface method for open tube interfaces}
The Stokes part of the equation is solved using the IIM-OT. This method requires the jump conditions for the solution to be known a priori. The velocity decomposition method is based on the fact that the jump conditions for the Navier-Stokes equations and the Stokes equations with the same interfacial forces are identical \cite{beale2009velocity}, which is true because the no-slip condition implies that 
  \begin{equation} 
  [\mathbf{u}]=0 
  \end{equation} 
  \noindent
  since the interface moves at the velocity of the local fluid \cite{lai2001remark}. 
  Taking the material derivative of velocity gives
  \begin{equation} 
  \left[\frac{\partial \mathbf{u}}{\partial t}+\nabla \mathbf{u}\cdot\mathbf{u}\right]=0.
  \end{equation}
  \noindent
 Therefore, the advection term does not contribute to the jump conditions \cite{lai2001remark}. 

 Since the jump conditions are known, the immersed interface method for open tube interfaces needs to only correct the incompressible Stokes equations. First, the incompressible Stokes equations are converted into a series of uncoupled Poisson problems. The Poisson problem for pressure, $p_s$ is found by taking the divergence of Eq. \ref{eq:stokepart1} and use the incompressibility condition
\begin{equation}\nabla\cdot(-\mu\Delta\mathbf{u_s})=-\nabla\cdot(\nabla p_s) +\nabla\cdot\mathbf{F}\end{equation}
\begin{equation}\Delta p_s =\nabla\cdot\mathbf{F}.\end{equation}
\noindent
 Once $p_s$ is known, Eq. \ref{eq:stokepart1} reduces to a Poisson problem for each component of velocity, $\mathbf{u}_s$. Therefore, to solve the Stokes part it is sufficient to use the immersed interface method for Poisson problems. 

The immersed interface method for Poisson problems was developed by  Ref.~ \cite{leveque1994immersed}. The Poisson problems are discretized using the finite difference method. 
 Therefore, to apply the IIM-OT to the Stokes equations, it is only necessary to compute the corrections terms for the first- and second-order spatial derivatives. Once the jump conditions for the open tube interface are known, the correction terms can be computed using the immersed interface method for Poisson problems.  

\subsection{Regular part: projection method and semi-Lagrangian method}
The regular part of the solution satisfies Eqs. \ref{eq:NSregular1}--\ref{eq:NSregular3}. The advection terms and the body force $\mathbf{F}_b$ are discretized using the backward difference formula, where the upstream values of velocity are found using the semi-Lagrangian method. A projection method is used to enforce the incompressibility condition.

\subsubsection{Semi-Lagrangian method}
The advection terms,
$\frac{\partial \mathbf{u}_r}{\partial t}+\mathbf{u}\cdot\nabla\mathbf{u}_r$ and $\frac{\partial \mathbf{u}_s}{\partial t}+\mathbf{u}\cdot\nabla\mathbf{u}_s$, can be thought of as material derivatives or the derivatives along the path of a fluid particle. In the semi-Lagrangian method, the material derivative is computed at Eulerian grid points, which combines the Lagrangian and Eulerian variable frameworks and has several benefits over a purely Lagrangian or Eulerian framework. 

The material derivative is computed at Eulerian grid points. Therefore, the resolution of the grid does not deteriorate or get distorted with the fluid flow, which can occur in Lagrangian methods. 

 Particles cannot physically cross the impermeable immersed interface so that the material derivatives will be smooth, which reduces the number of correction terms needed in the immersed interface method.   For example,  correction terms would be needed for the time derivatives computed on an Eulerian grid when the interface crosses over a grid point in a time step.  
Additionally, $\nabla \mathbf{u}$ is discontinuous at $\Gamma$, so the terms,
$\mathbf{u}\cdot\nabla\mathbf{u}_r$ and $\mathbf{u}\cdot\nabla\mathbf{u}_s$, would need to be corrected near the interface if  discretized on the Eulerian grid \cite{layton2003semi-Lagrangian,xiu2001semi}. 
Additionally, the nonlinear terms of the Navier-Stokes equations do not need to be discretized explicitly in the semi-Lagrangian method.

The semi-Lagrangian method is used to compute the velocity along the path of a fluid particle that passes through the grid point $\mathbf{x^{n+1}}$ at time $t_{n+1}$, which is used to compute the material derivative of the velocity of the particles.  This is done by first computing the upstream positions of the particle,  $\mathbf{x^{n}}$ and $\mathbf{x^{n-1}}$, at two previous times, $t_n$ and $t_{n-1}$, respectively.   An example of a 1D particle path is shown in Fig. \ref{fig:particlePath}.

The upstream position of the particle located on the Eulerian fluid grid point, $\mathbf{x}_0$, at time $t_{n+1}$ can be found by integrating 
\begin{equation}
    \frac{d\mathbf{x}(t)}{dt}=\mathbf{u}(\mathbf{x}(t),t), \quad \mathbf{x}(t_{n+1})=\mathbf{x}_0,
\end{equation}
backwards in time. The predictor-corrector method 
\begin{align}
    \mathbf{x}^*&=x_0-\frac{\Delta t}{2}\mathbf{u}\left(\mathbf{x}_0-\frac{\Delta t}{2}\mathbf{u}^{n+\frac12},t_{n+\frac12}\right)\\
    \mathbf{x}^n&=\mathbf{x}_0-\Delta t \mathbf{u}(\mathbf{x}^*,t_{n+\frac12})
    \end{align}
    \noindent
and
\begin{align}
    \mathbf{x}^*&=x_0-\Delta t\mathbf{u}\left(\mathbf{x}_0-\Delta t\mathbf{u}^{n},t_n\right)\\
    \mathbf{x}^{n-1}&=\mathbf{x}_0-2\Delta t \mathbf{u}(\mathbf{x}^*,t_n)
    \end{align}
    \noindent
    can be used to find the upstream positions at two previous time steps with second-order temporal accuracy. Notice that $\mathbf{u}^{n+\frac12}$ is approximated using the time extrapolation $\frac32\mathbf{u}^n-\frac12\mathbf{u}^{n-1}$. 
    
It is unlikely that the upstream positions coincide with grid points. Therefore the velocity at upstream values
\begin{align}
    \mathbf{\tilde{u}}_r^n&=\mathbf{u}_r(\mathbf{x}^n,t_n)\\
       \mathbf{\tilde{u}}_r^{n-1}&=\mathbf{u}_r(\mathbf{x}^{n-1},t_{n-1})
\end{align}
\noindent
are interpolated in space from the velocity at grid points using cubic Lagrangian interpolation. This interpolation has 4th order accuracy in space. Once the upstream values of velocity are known, the material derivative is found using the second-order backwards difference formula. Therefore, Eq. \ref{eq:NSregular1} can be discretized using 
\begin{equation}
    \frac{3\mathbf{u}_r^{n+1}-4\mathbf{\tilde{u}}_r^n+\mathbf{\tilde{u}}_r^{n-1}}{2\Delta t}+\nabla p_r^{n}=\mu\Delta\mathbf{u}^{n+1}_r+\mathbf{F}^{n+1}_b.
\end{equation}

\subsubsection{Projection method}

The coupling of the velocity components and pressure in the Navier-Stokes equations would create a large finite difference system if all components were solved simultaneously. Instead, the projection method is used to solve the regular part. In this procedure, an intermediate value of velocity ${\bf u}^*$ and pressure $p^*$ are found that satisfy 
 \begin{equation}
 \frac{D {\bf u^*}}{D t}=-\nabla p^*+\mu\nabla^2{\bf u^*}+{\bf F}.
 \end{equation}
 \noindent
The intermediate values are then projected, $P({\bf u}^*)={\bf u}$, into the subspace of divergence-free vector fields so that $ \nabla\cdot {\bf u}=0.$ This works because every vector field  $\mathbf{w}$ can be written as a Hodge decomposition 
$\mathbf{w}=\mathbf{v}+\nabla\phi$ where $\mathbf{v}$ is divergence-free. 
This projected value $\mathbf{u}$ satisfies the Navier-Stokes equations.

The regular part of the Navier-Stokes equations can be numerically solved using the following projection method.  An intermediate velocity value, ${\bf u}^*$ is found by solving
\begin{equation}
    \frac{3\mathbf{u}_r^*-4\mathbf{\tilde{u}}_r^n+\mathbf{\tilde{u}}_r^{n-1}}{2\Delta t}+\nabla p_r^{n}=\mu\Delta\mathbf{u}^{*}_r+\mathbf{F}^{n+1}_b.
\end{equation}
\noindent
where $p_r^n$ is the value of pressure from the previous time step. The intermediate value $\mathbf{u}^*_r$ is approximately projected into divergence-free space by defining $\phi$ as
\begin{equation}
    \mathbf{u}_r^{n+1}=\mathbf{u}^*_r-(\Delta t)\nabla \phi^{n+1},
\end{equation}
\noindent
which can be found by solving
\begin{align}
\Delta\phi^{n+1}&=\frac{1}{\Delta t}\nabla\cdot{\bf u}_r^*.
\end{align}
\noindent
Then, velocity and the gradient of pressure can be found by
\begin{equation} 
{\bf u}_r^{n+1}={\bf u}_r^*-\Delta t \nabla\phi^{n+1}
\end{equation}
\begin{equation}
\nabla p_r^{n+1}=\nabla p^{n}+\frac32 \nabla \phi^{n+1}-\mu \Delta t \nabla^3\phi^{n+1}.
\end{equation}

\subsection{Time-stepping}
The interface movement is govern by Eq. \ref{eq:noSlip}. 
Suppose the interface position at time $t_n=n\Delta t$ is represented by $N_b$ boundary markers $\mathbf{X}^{n}=\{\mathbf{X}_i^{n}\}_{i=1}^{N_b}$. The velocity at each marker $\mathbf{U}^{n}=\{\mathbf{U}_i^{n}\}_{i=1}^{N_b}$ can be found by interpolating the velocity from the fluid grid to the markers using a second-order method such as bi-linear interpolation. Care must be taken during this interpolation since the velocity near the interface is not smooth. The velocity could be corrected near the interface or a one-sided approximation for velocity could be used where all interpolation points are on the same side of the interface. 

For the time stepping to be second-order, the velocity on the interface at two previous time levels,  $\mathbf{U}^{n}$ and  $\mathbf{U}^{n-1}$, are needed. The interface position is then updated as follows:

\begin{equation}
    \mathbf{X}^{n+1}=\mathbf{X}^n+\Delta t\left( \frac{3}{2} \mathbf{U}^n -\frac{1}{2} \mathbf{U}^{n-1} \right).
    \label{eq:IntUpdate}
\end{equation}
To move the interface initially, forward Euler's method was used.

\subsection{Velocity decomposition method algorithm outline}

At time $t_n$, the following are known: $\mathbf{u}^i$, $p^i$, $\mathbf{u}_s^i$, $p_s^i,$   $\mathbf{u}_r^i$, $p_r^i,$ and ${\bf X}^i$ for $i=n$ and $i=n-1$. The following procedure can be used to compute the solution at $t_{n+1}$.
\begin{enumerate}
\item Update interface position $\mathbf{X}^{n+1}$.
\begin{enumerate}
\item [a)] Interpolate velocity at the interface $\mathbf{U}^i$ from fluid grid $\mathbf{u}^i$ using
\begin{equation}
    \mathbf{U}^n\approx \mathbf{u}^n(\mathbf{X}^n,t_n)
\end{equation}
\item [b)] The interface positions ${\bf X}^{n+1}$ is updated using the no slip condition
\begin{equation}
    \mathbf{X}^{n+1}=\mathbf{X}^n+\Delta t\left( \frac{3}{2} \mathbf{U}^n -\frac{1}{2} \mathbf{U}^{n-1} \right).
\end{equation}
\item[c)] Compute the interfacial force $\mathbf{f}^{n+1}$ from the interface position $\mathbf{X}^{n+1}$.
\end{enumerate}
\item Compute the Stokes solution, $\mathbf{u}^{n+1}_s$ and $p^{n+1}_s$.
\begin{enumerate}
\item [a)]  Use force $\mathbf{f}^{n+1}$ to find the correction terms, $\mathbf{C}^{n+1}_{p_h}$ and $\mathbf{C}^{n+1}_{\mathbf{u}_h}$, for pressure and velocity respectively using the immersed interface method.
    \item[b)] Solve the following Poisson problem for pressure,  $p^{n+1}$:
    \[\Delta_h p^{n+1}=\mathbf{C}^{n+1}_{p_h} \]
    \item[c)] Compute the discrete pressure gradient, $\nabla_h p^{n+1}$ using correction terms. 
    \item[d)] Solve the following Poisson problems for each component of velocity,  $\mathbf{u}^{n+1}$.
    \[\Delta_h \mathbf{u}^{n+1}=-\nabla_h p^{n+1}+\mathbf{C}^{n+1}_{\mathbf{u}_h} \]
\end{enumerate}
\item Compute regular solution, $\mathbf{u}^{n+1}_r$ and $p^{n+1}_r$. 
\begin{enumerate}
\item [a)] Use semi-Lagrangian method to compute $F_b=-\frac{d {\bf u}_s}{dt}$, $\mathbf{\tilde{u}}_r^{n}$, and $\mathbf{\tilde{u}}_r^{n-1}.$
\item [b)] Solve momentum equation for intermediate regular velocity, $\mathbf{u}^*$.
\item [c)] Compute $\phi^{n+1}$ needed to project into divergent-free space. 
\item [d)] Find $\mathbf{u}^{n+1}_r$ and $p_r^{n+1}$ using $\phi^{n+1}$. 
\end{enumerate}
\item Compute the full solution at time $t_{n+1}$ by adding the Stokes and regular part. 
\end{enumerate}
Forward Euler's method is used for the first step of the time stepping methods. The local truncation error is $O(\Delta t^2)$ and this is only used for one time step, second-order temporal accuracy is still maintained.

\section{Numerical results}

\subsection{Stokes Results}

A grid refinement study was conducted to test the spatial convergence rate for the immersed interface method in an open tube for the Stokes equations in 3D axisymmetric cylindrical coordinates shown in Eqs. \ref{eq:StokesCyl1} -- \ref{eq:StokesCyl3}. In this test, the tube was displaced and allowed to return to its resting position as a tube with a constant radius. In addition to confirming the order of the method, this test also aims to validate that approximate 3D Poiseuille flow is achieved when the tube has near constant radius. 

\begin{equation} \frac{\partial p}{\partial r}=\mu\left(   \frac{1}{r}\frac{\partial }{\partial r}\left(r \frac{\partial w}{\partial r}\right)+\frac{\partial^2 w}{\partial z^2}-\frac{w}{r^2}\right)
\label{eq:StokesCyl1}
\end{equation}
\begin{equation}
\frac{\partial p}{\partial z}=\mu\left(   \frac{1}{r}\frac{\partial }{\partial r}\left(r \frac{\partial u}{\partial r}\right)+\frac{\partial^2 u}{\partial z^2}\right)
\label{eq:StokesCyl2}
\end{equation}
\begin{equation}
\frac{1}{r}\frac{\partial (r u)}{\partial r}+\frac{\partial w}{\partial z}=0
\label{eq:StokesCyl3}
\end{equation}

  For the cylindrical coordinate test cases, the domain is assumed to be radially symmetric. This allows three dimensional kinetics with two dimensional computational costs. Although, the fluid domain is  $\Omega=[-1.625\text{cm},1.625\text{cm}]\times[0,2\pi]\times[0\text{cm},4.25\text{cm}]$, the solution is only computed on the computational domain $\Omega=[0\text{cm},1.625\text{cm}]\times[0\text{cm},4.25\text{cm}]$ which is the $\theta=0$ slice of the fluid domain. $R=.7$ cm is the resting radius of the tube.
  
  

For this case, the interface was initially displaced in the $r$ direction according to the function
\begin{equation} r(z)=.2e^{-(z-\frac{L}{2})^2}-.2e^{-(\frac{L}{2})^2}+R\end{equation}
where $R$ is the resting radius of the tube. The interface was then allowed to return to its resting position as a straight tube.

Flow is driven through the tube by prescribing a pressure gradient over the computational domain using Dirichlet boundary conditions.
At the inlet, it is assumed that $w$ has a parabolic profile consistent with 3D Poiseuille flow.  In 3D axisymmetric cylindrical coordinates, the pressure gradient and velocity are related by
\begin{equation}w=-\frac{dp}{dz}\frac{1}{4\mu}(R^2-y^2)\end{equation}
\begin{equation}u=-\frac{dp}{dr}=0.\end{equation}

\begin{itemize}
\item Inlet Boundary ($z=0$)
\[p|_{z=0}=\begin{cases} 
      0 & |r|< R \\
     \frac{ 8\mu\bar{w}L}{R^2} &  |r|\leq R 
   \end{cases},\quad\quad w|_{z=0}=\begin{cases} 
      0 & r< -R \\
     \frac{ 2\bar{w}(R^2-r^2)}{R^2} & | r|\leq R \\
      0 & R< r
   \end{cases},\quad\quad u|_{z=0}=0\]
 
\item Outlet Boundary ($z=L$)
\[p|_{z=L}=0, \quad\quad \frac{\partial w}{\partial z}|_{z=L}=0,
\quad\quad\frac{\partial u}{\partial z}|_{z=L}=0\]
\item Top Boundary ($r=H$)
\[\frac{\partial p}{\partial r}|_{r=H}=0,
\quad\quad \frac{\partial w}{\partial r}|_{r=H}=0,\quad\quad \frac{\partial u}{\partial r}|_{r=H}=0\]
\item Bottom Boundary ($r=0$)
\[\frac{\partial p}{\partial r}|_{r=0}=0,\quad\quad \frac{\partial w}{\partial r}|_{r=0}=0,\quad\quad u|_{r=0}=0\]
\end{itemize}
The remainder of the parameters used for this experiment are shown in Table \ref{tab:StokeCylParm}. 

\begin{table}[ht!]
\caption{Parameters used for the spatial convergence study of the IIM-OT for 3D Stokes flow in axisymmetric cylindrical coordinates.}  
\label{tab:StokeCylParm}
\begin{center}
\begin{tabular}{c|c|c}
Parameter &Symbol&Value\\\hline
Viscosity&$\mu$&0.1 gm/(cm$\cdot$s)\\
Density&$\rho$&1 gm/cm$^3$\\
Domain  length&$L$&4.25 cm\\
Domain height& $H$&1.625 cm\\
Initial tube radius&$R$&0.7 cm\\
Interface control points& $N_{b}$ & 100\\
Simulation length&$T$& 4 s\\
Time steps &$N_t$& 2000\\
Time step size & $\Delta t$ & 2e-3 s \\
Tether force constant &$a_{Tether}$&10 gm/s$^2$ \\
Elastic force constant &$a_{Elastic}$&.1 gm/s$^2$ \\
Average Inlet Velocity& $\bar{w}$ & 1 cm/s
\end{tabular}
\end{center}
\end{table}

The solution was computed at various resolutions and then compared to a $1089\times417$ high-resolution solution. The results of the convergence study shown in Table \ref{tab:StokeCylSpaceConvResults} indicate that the method is second-order accurate in space. 


\begin{table}[ht!]
\caption[Spatial convergence results of the IIM-OT for 3D Stokes flow in axisymmetric cylindrical coordinates.]{Spatial convergence results of the IIM-OT for Stokes flow in cylindrical coordinates. Solutions are compared to a high resolution, $1089\times417$ solution at time $t=0$.}
\label{tab:StokeCylSpaceConvResults}
\begin{center}
\begin{tabular}{cc|cc|cc|cc}
\multicolumn{2}{c}{Grid Size} &  \multicolumn{2}{c}{p} & \multicolumn{2}{c}{w}& \multicolumn{2}{c}{u}\\
$N_z$ & $N_r$  & $||\cdot||_{\infty}$ & Order & $||\cdot||_{\infty}$ & Order& $||\cdot||_{\infty}$ & Order  \\\hline
35 &14 &0.03741&    &    0.25162&    &     0.10789    &\\
69& 27 &0.00796&    2.23&    0.10349&    1.28&    0.02643    &2.03\\
137& 53 &0.00189&    2.07&        0.01862&    2.47&    0.00628    &2.07\\
273& 105 &0.00044&    2.12&    0.00456&    2.03&    0.00145    &2.12\\
545&209&0.00008&2.37&        0.00027&4.07&        0.00029 &2.34\\
\end{tabular}
\end{center}
\end{table}

Fig. \ref{fig:StokeCylSlope} shows the interface  position overlaid on a quiver plot that show the velocity as an arrow with components $(w,u)$ various times. Since the most dramatic changes occur in the beginning of the simulation, the Fig. \ref{fig:StokeCylSlope0}, \ref{fig:StokeCylSlope2}, and \ref{fig:StokeCylSlope3} show the results at $t=0$, $t=.05s$, and $t=.1s$ respectively. Recall that because of the no-slip condition on the interface, the interface moves at the local fluid velocity.  Therefore, the velocity slope field in Fig. \ref{fig:StokeCylSlope0} indicates that the interface is moving towards its resting position as a straight tube (shown in Fig. \ref{fig:StokeCylSlope4}).
Initially in Fig. \ref{fig:StokeCylSlope0}, some of the fluid is pushed to the left towards the proximal end of the tube as the interface relaxes. As the interface gets closer to its resting configuration, it exerts less force on the fluid. Therefore, in Figs. \ref{fig:StokeCylSlope2}--\ref{fig:StokeCylSlope4}, all fluid in the tube moves to the right towards the distal end of the tube.  Fig. \ref{fig:StokeCylSlope4} shows the resting position of the interface, in which the velocity exhibits a parabolic profile inside the interface and dissipates to zero outside of the interface. 

\begin{figure}[htp!]
    \centering
    \begin{subfigure}[b]{0.7\textwidth}
        \includegraphics[width=\textwidth]{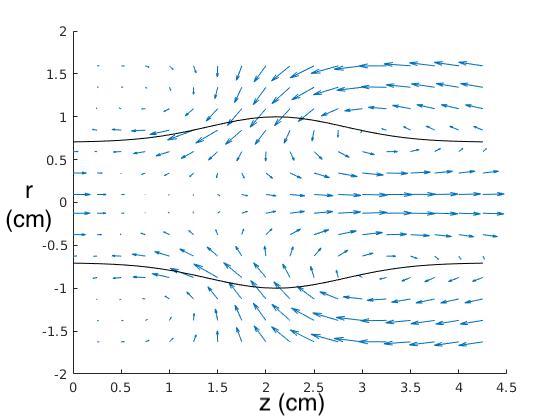}
        \caption{The interface position and velocity quiver plot at $t=0$ s shows the interface is moving towards $r=\pm0.7$ cm. }
        \label{fig:StokeCylSlope0}
    \end{subfigure}
    ~
    \begin{subfigure}[b]{0.7\textwidth}
        \includegraphics[width=\textwidth]{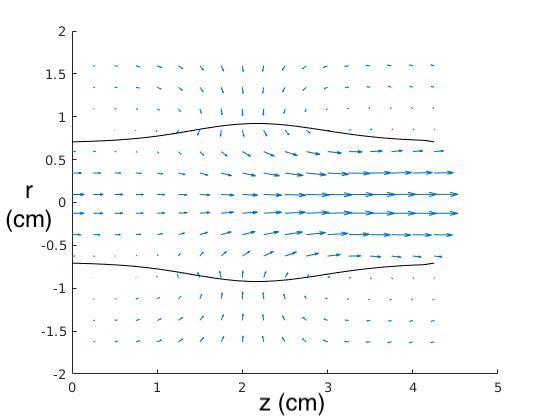}
        \caption{The interface position and velocity quiver plot at $t=0.05$ s shows the interface is moving towards $r=\pm0.7$ cm.}
        \label{fig:StokeCylSlope2}
    \end{subfigure}
        \caption[The interface is shown at various times as it relaxes from its displaced in the axisymmetric cylindrical computational domain.]{The interface (black line) is shown at various times as it relaxes from its displaced in the axisymmetric cylindrical computational domain. The interface position is superimposed on a quiver plot that show the velocity as an arrow with components $(w,u)$.}
    \label{fig:StokeCylSlopea}
\end{figure}
    \begin{figure}[htp!]\ContinuedFloat
    \centering
    \begin{subfigure}[b]{0.7\textwidth}
        \includegraphics[width=\textwidth]{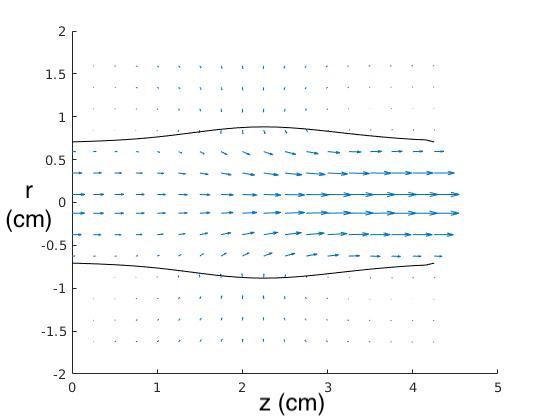}
        \caption{The interface position and velocity quiver plot at $t=0.1$ s shows the interface is moving towards $r=\pm0.7$ cm.s}
        \label{fig:StokeCylSlope3}
    \end{subfigure}
    ~
    \begin{subfigure}[b]{0.7\textwidth}
        \includegraphics[width=\textwidth]{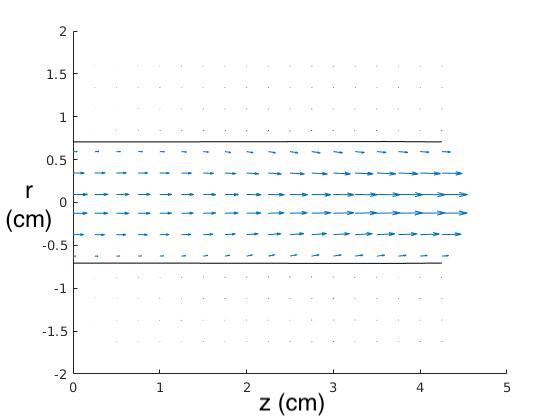}
        \caption{The interface position and velocity quiver plot at $t=2$ s shows the interface at its resting position where $r=\pm0.7$ cm.}
        \label{fig:StokeCylSlope4}
    \end{subfigure}
    \caption[The interface is shown at various times as it relaxes from its displaced in the axisymmetric cylindrical computational domain.]{The interface (black line) is shown at various times as it relaxes from its displaced in the axisymmetric cylindrical computational domain. The interface position is superimposed on a quiver plot that show the velocity as an arrow with components $(w,u)$.}
    \label{fig:StokeCylSlope}
\end{figure}

 \begin{figure}[htp!]
    \centering
    \begin{subfigure}[b]{0.7\textwidth}
        \includegraphics[width=\textwidth]{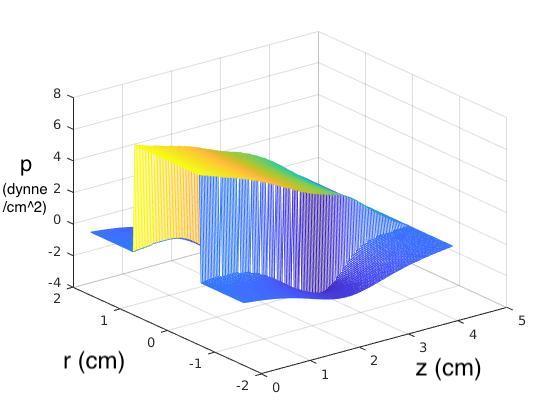}
        \caption{The initial pressure in the computational domain at $t=0$s.}
        \label{fig:StokeCylPProf0}
    \end{subfigure}
    ~
    \begin{subfigure}[b]{0.7\textwidth}
        \includegraphics[width=\textwidth]{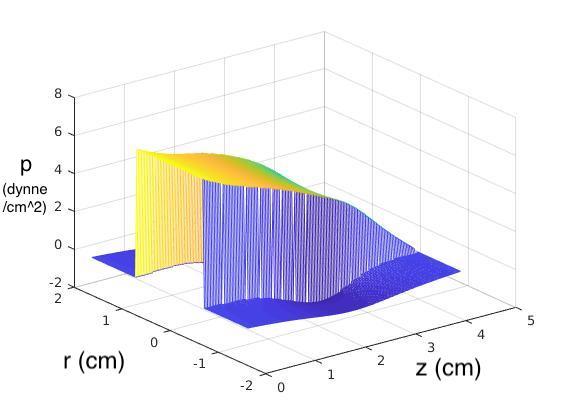}
        \caption{The pressure at $t=0.05$ s as the interface is returning to its resting position. }
        \label{fig:StokeCylPProf0_5}
    \end{subfigure}
        \caption{Pressure in the axisymmetric cylindrical computational domain at various times as the interface relaxes from its displaced position to rest.}
    \label{fig:StokeCylPProfb}
\end{figure}
    \begin{figure}[htp!]\ContinuedFloat
    \centering
    \begin{subfigure}[b]{0.7\textwidth}
        \includegraphics[width=\textwidth]{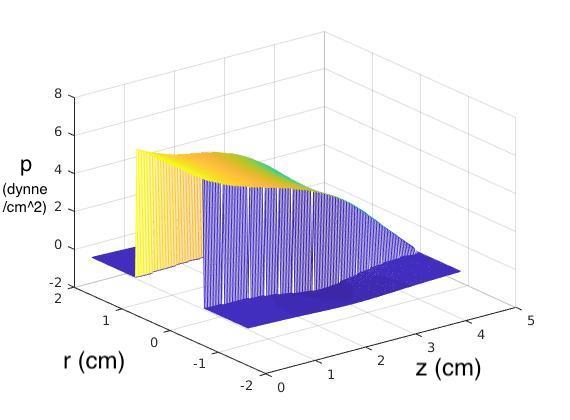}
        \caption{The pressure at $t=0.1$ s as the interface is returning to its resting position. }
        \label{fig:StokeCylPProf1}
    \end{subfigure}
    ~
    \begin{subfigure}[b]{0.7\textwidth}
        \includegraphics[width=\textwidth]{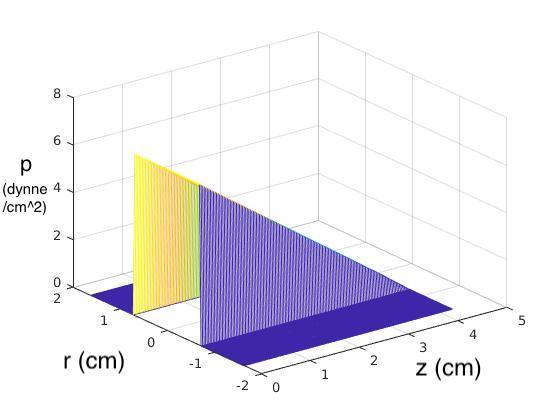}
        \caption{The pressure is linearly decreasing inside the tube and zero outside at $t=2$s when the interface is in its resting position. }
        \label{fig:StokeCylPProf1_5}
    \end{subfigure}
    \caption{Pressure in the axisymmetric cylindrical computational domain at various times as the interface relaxes from its displaced position to rest.}
    \label{fig:StokeCylPProf}
\end{figure}

 The pressure in the computational domain at $t=0$s, $t=.05$s, $t=.1$s and $t=2$s are shown in Fig. \ref{fig:StokeCylPProf}.
 The IIM-OT captures the discontinuity in pressure across the interface as can be seen in Fig. \ref{fig:StokeCylPProf}. Additionally, pressure decreases along the length of the domain inside of the interface, which creates a pressure gradient that drives fluid through the open tube. Another pressure gradient outside of the tube can be seen in Fig. \ref{fig:StokeCylPProf0} which drives the fluid to restore the interface to its resting position. Finally, Fig. \ref{fig:StokeCylPProf1_5} shows a linearly decreasing pressure that is typical of Poiseuille flow in a cylinder.

\subsection{Navier-Stokes Equations}

\subsubsection{Initial solutions}
For the velocity decomposition method, an initial solution is needed for both the full solution and the regular part. In the two simulations given in Section \ref{sec:NSCylResults}, the interface is initially a straight tube with a constant radius. Therefore, the initial solution of the full solution is taken to be approximately Poiseuille flow. Since the interfacial force is in the Stokes part, the Stokes part will approximate Poiseuille flow. Therefore, the Stokes part and the full solution will have identical initial solutions. The initial solution of the regular part will be identically zero. The initial solutions are 
\begin{equation}
    p(r,z,0)=p_s(r,z,0)=\begin{cases} 
      0 & |r|>R \\
     \frac{ 8\mu\bar{w}(L-z)}{R^2} &  |r|\leq R 
   \end{cases}
\end{equation}
\begin{equation}
    u(r,z,0)=u_s(r,z,0)=0
\end{equation}
\begin{equation}
    w(r,z,0)=w_s(r,z,0)=\begin{cases} 
      0 & |r|>R \\
     \frac{ 2\bar{w}(R^2-r^2)}{R^2} & | r|\leq R \\
      \end{cases}
\end{equation}
\begin{equation}
    p_r(r,z,0)=u_r(r,z,0)=w_r(r,z,0)=0,
\end{equation}
\noindent
where $\bar{w}$ is the average inlet velocity.

\subsubsection{Spatial convergence study: tethers with sinusoidal movement}
\begin{figure}[htp!]
    \centering
    \includegraphics[height=3in]{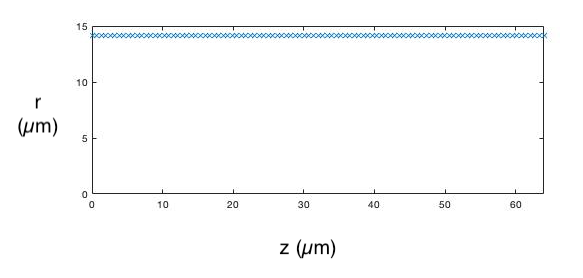}
    \caption{Interface tether position at $t=1$ ms.}
    \label{fig:tetherPositionNSCylSpatConv}
\end{figure}

A grid refinement study was conducted to test the spatial convergence rate for the IIM-OT for the Navier-Stokes equations in 3D axisymmetric cylindrical coordinates.  In this simulation, the initially straight interface is actively moved by displacing the tether anchor positions in time. The $r$-position of the tether anchor at position $z$ at time $t$  is given by
\begin{equation} R_{tether}(z,t)=R\left(1-\alpha\sin\left(\frac{2\pi t}{\beta}\right)\sin\left(\frac{2\pi z}{L}\right)\right)
       \end{equation}
\noindent 
where $R$ is the initial interface radius, the period is $\beta=10 s$ and the amplitude is $\alpha=0.1$ $\mu$m. The positions of the $N_b=100$ interface tethers at $t=1$ ms are shown in Fig.  \ref{fig:tetherPositionNSCylSpatConv}.

Even though the fluid domain is  $\Omega=[-32\mu m,32 \mu m]\times[0,2\pi]\times[0\mu m,64 \mu m]$, the solution is only computed on the computational domain $\Omega=[0 \mu m,32 \mu m]\times[0\mu m,64\mu m]$ which is the $\theta=0$ slice of the fluid domain.  The initial radius of the tube is $R=14.14$ $\mu$m.   

The boundary conditions and initial solutions for the full solution and the Stokes and regular parts are given in Section \ref{subsec:NScylBC}.  The parameters for this simulation are shown in Table \ref{tab:NSCylSpatialParms}.

\begin{table}[ht!]
\caption{Parameters used for the spatial convergence study of the IIM-OT for 3D Navier-Stokes flow in axisymmetric cylindrical coordinates. }
\label{tab:NSCylSpatialParms}
\begin{center}
\begin{tabular}{|c|c|c|}\hline
Parameter &Symbol&Value\\\hline\hline
Viscosity&$\mu$& 0.0175 gm/(cm$\cdot$s)\\
Density&$\rho$& 1.055 gm/cm$^3$\\
Domain  length&$L$&64 $\mu$m\\
Domain height& $H$&32 $\mu$m\\
Initial tube radius&$R$& 14.14 $\mu$m\\
Simulation length&$T$&1e-3 s\\
Number of time steps &$N_t$&16 \\
Time step size &$\Delta t$&6.25e-5 s \\
Tether force constant & $a_{Tether}$& 2.5e-2 gm/s$^2$\\
Elastic force constant & $a_{Elastic}$ &2.5e-3 gm/s$^2$\\
Average inlet velocity &$\bar{w}$& 1 cm/s \\\hline
\end{tabular}
\end{center}
\end{table}

In order to test the spatial convergence of the method, the solution is computed at various spatial resolutions while the time step size $\Delta t$ is held fixed. The solutions are compared with a  $513\times 1025$ high-resolution solutions during the at time $t=1$ ms.  The results in Table \ref{tab:NSCylSpatConvResults} indicate that the method converges with second-order spatial accuracy for all the fluid variables. Additionally, the position $X$ of the interface $\Gamma$ also converges with second-order accuracy.

\begin{table}[ht!]
\caption[Spatial convergence results for IIM-OT for 3D Navier-Stokes flow in axisymmetric cylindrical coordinates.]{Spatial convergence results for IIM-OT for Navier-Stokes flow in cylindrical coordinates compared to a high resolution $513\times1025$ solution taken at time $t=1$ms. }
\label{tab:NSCylSpatConvResults}
\begin{center}
\begin{tabular}{cc|cc|cc|cc|cc}
\multicolumn{2}{c}{Grid Size} &  \multicolumn{2}{c}{$p$}
&\multicolumn{2}{c}{$w$}& \multicolumn{2}{c}{$u$}& \multicolumn{2}{c}{$X$}\\
$N_r$ &$N_z$ & $||\cdot||_{\infty}$ & Order & $||\cdot||_{\infty}$ & Order& $||\cdot||_{\infty}$ & Order  & $||\cdot||_{\infty}$ & Order  \\\hline
&&1.0e-05 $\times$&&&&&&1.0e-03 $\times$&\\                    
33&65&0.1929&&        0.3530    &&    0.1358    &&0.3365&\\
65&129&0.0259&    2.90&    0.0410&    3.11&    0.0349&    1.96&0.0365&    3.20\\
129&257&0.0064&    2.02&    0.0110&    1.90&    0.0131&    1.41&0.0085&    2.10\\
257&513&0.0008&    3.00&    0.0026&    2.08&    0.0028&    2.23&0.0016&    2.41    \\\hline
\end{tabular}
\end{center}
\end{table}

\subsubsection{Temporal convergence study: moving tethers}
In order to test temporal convergence, a series of simulations are run in which an initially a straight tube is actively moved by displacing the time-dependent tether anchors. This test also shows the ability of the method to sharply capture the discontinuities of the fluid solutions and the movement of an active tube.  Although, the fluid domain is  $\Omega=[0{\rm  \mu m},32{\rm  \mu m}]\times[0,2\pi]\times[0{\rm  \mu m},64 {\rm  \mu m}]$, the solution is only computed on the computational domain $\Omega=[0{\rm  \mu m},32{\rm  \mu m}]\times[0{\rm  \mu m},64{\rm  \mu m}]$ which is the $\theta=0$ slice  of the fluid domain. $R=14.1$ $\mu$m is the initial radius of the tube.

\begin{figure}[htp!]
    \centering
    \includegraphics[height=2.5in]{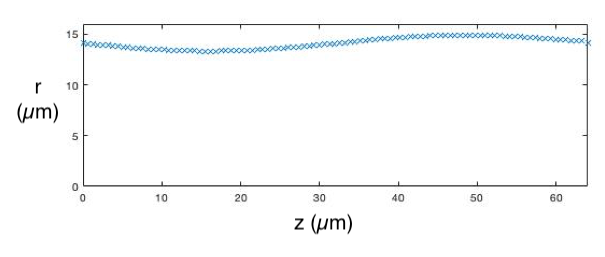}
    \caption{Interface tether anchor position at $t=0.1$ s.}
    \label{fig:tetherPositionNSCyltempConv}
\end{figure}

The $r$-position of the tethers  at position $z$ at time $t$ for the interface is given by
\begin{equation} R_{tether}(z,t)=R\left(1-\alpha\sin\left(\frac{\pi t}{\beta}\right)\sin\left(\frac{2\pi z}{L}\right)\right)
       \end{equation}
       \noindent
       where $R$ is the initial interface radius, the period is $\beta=.2$ s and the amplitude is $\alpha=0.1$ $\mu$m. The positions of the $N_b=100$ interface tethers anchors at $t=0.1$ s are shown in Fig. \ref{fig:tetherPositionNSCyltempConv}. 
The boundary conditions and initial solution are given in Section \ref{subsec:NScylBC}. The parameter values for this simulation are given in Table \ref{tab:NSCylTempConvParms}.

\begin{table}[ht!]
\caption{Parameters used for the temporal convergence study of the IIM-OT for 3D Navier-Stokes flow in axisymmetric cylindrical coordinates }
\label{tab:NSCylTempConvParms}
\begin{center}
\begin{tabular}{|c|c|c|}\hline
Parameter &Symbol&Value\\\hline\hline
Viscosity&$\mu$& .0175 gm/(cm$\cdot$s)\\
Density&$\rho$&1.055 gm/cm$^3$\\
Domain  length&$L$& 64 $\mu$m\\
Domain height& $H$&32 $\mu$m\\
Initial tube radius&$R$& 14.1 $\mu$m\\
Simulation length&$T$&.1 s\\
Grid points in z &$N_z$& 257 \\
Grid points in r &$N_r$& 129 \\
Interface control points& $N_b$ & 100\\
Tether force strength &$a_{\rm tether}$&2.5e-4 gm/s$^2$\\
Elastic force strength &$a_{\rm elastic}$&2.5e-4 gm/s$^2$\\
Average inlet velocity &$\bar{w}$& 10 $\mu$m/s \\\hline
\end{tabular}
\end{center}
\end{table}

In order to determine the temporal accuracy, the solution was computed with various sized time steps while the spatial resolution remained fixed. The solutions were then compared to a high-resolution solution computed with $N_t=512$ time steps with a step size of $\Delta t= .0002$ s.
 The results for this study are given in Table \ref{tab:NSCylTempConvResults}. This simulation indicates that the method converges with second-order temporal accuracy for velocity and interface position and near second-order temporal accuracy for pressure. 

\begin{table}[ht!]
\caption{Temporal convergence results for IIM-OT for 3D Navier-Stokes flow in axisymmetric cylindrical coordinates.}
\label{tab:NSCylTempConvResults}
\begin{center}
\begin{tabular}{cc|cc|cc|cc|cc}
\multicolumn{2}{c}{Time Step} &  \multicolumn{2}{c}{$p$} &\multicolumn{2}{c}{$w$}& \multicolumn{2}{c}{$u$}& \multicolumn{2}{c}{$X$}\\
$N_t$ &$\Delta t$ & $||\cdot||_{\infty}$ & Order & $||\cdot||_{\infty}$ & Order& $||\cdot||_{\infty}$ & Order & $||\cdot||_{\infty}$ & Order  \\\hline
&&1.0e-07 $\times$&&&& & &1.0e-02$\times$&  \\                
32&.0031&0.7660&               &    0.0147&           &     0.0077    &&.2925&\\
64&.0016&0.2282&    1.75&    0.0035&    2.07&    0.0019&    2.02&.0730&2.00\\
128&.0008&0.0667&    1.77&    0.0009&    1.96&    0.0005&    1.93&.0177&2.04\\
256&.0004&0.0287&    1.22&    0.0002&    2.17&    0.0001&    2.32&.0038&2.21  \\
\end{tabular}
\end{center}
\end{table}

\begin{figure}[htp!]
    \centering
    \includegraphics[height=3in]{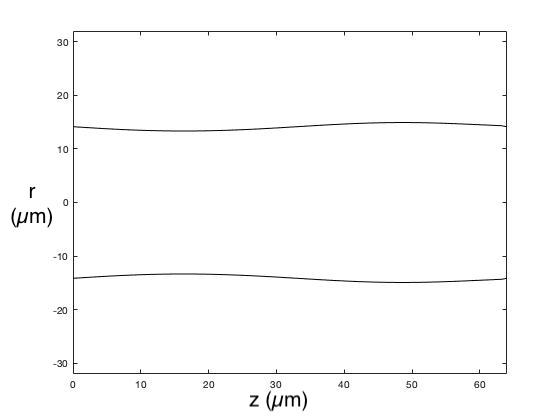}
    \caption{The interface position at time $t=.1$ s. The lower half of the domain is computed by imposing symmetry.  }
    \label{fig:NScylTempIntPosition}
\end{figure}

The interface position at time $t=0.1$s is shown in Fig. \ref{fig:NScylTempIntPosition}. The interface is also reflected into the domain $[-H,0]\times [0,L]$ by imposing symmetry over $r=0.$ It can be seen that moving the interface tethers is an effective way to generate interface movement.

\begin{figure}[!htp]
    \begin{subfigure}[b]{1\textwidth}
        \begin{center}
    \includegraphics[height=3.25in]{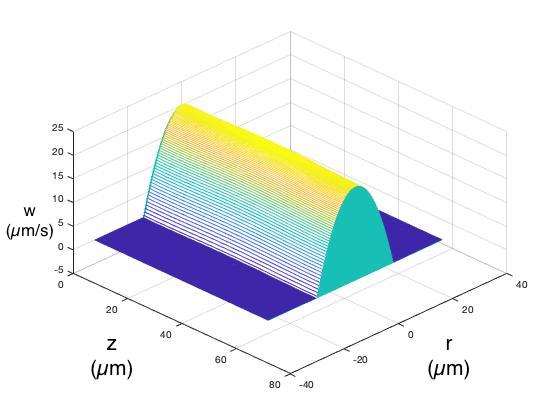}
     \end{center}
    \caption{The $w$ component of velocity at time $t=0$ s exhibits a parabolic profile inside the tube when the interface is straight. }
    \end{subfigure}
    \begin{subfigure}[b]{1\textwidth}
        \begin{center}
    \includegraphics[height=3.25in]{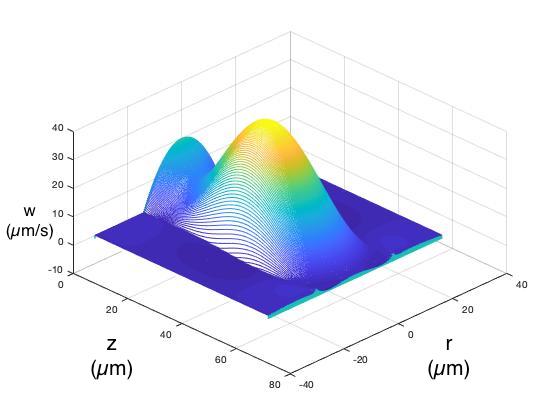}
     \end{center}
    \caption{The $w$ component of velocity at time $t=0.1$ s is greatly affected by the slight change in the interface position caused by the moving tether anchor points.}
    \end{subfigure}
    \caption[The $w$ component of velocity at time $t=0$ and at $t=0.1$ s are shown in the computational domain.]{The $w$ component of velocity at time (a) $t=0$ and at (b) $t=0.1$s  are shown in the computational domain and its reflection over $r=0$.   The value of $w$ in the lower half of the domain is computed  by imposing symmetry.}
    \label{fig:NScylW}
\end{figure}

\begin{figure}[!htp]
    \begin{subfigure}[b]{1\textwidth}
    \begin{center}
    \includegraphics[height=3.25in]{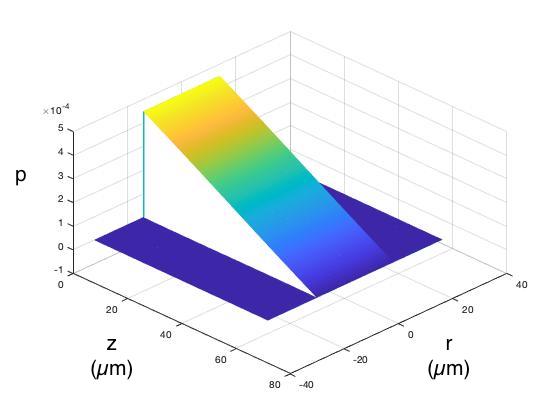}
    \end{center}
    \caption{The pressure $p$ at time $t=0$ in the computational domain is linearly decreasing inside the straight tube. }
        \end{subfigure}
    \begin{subfigure}[b]{1\textwidth}
        \begin{center}
    \includegraphics[height=3.25in]{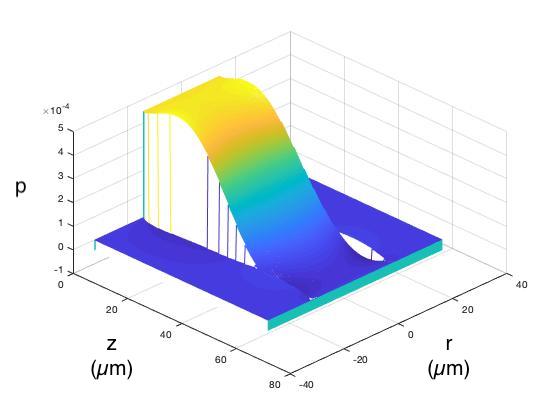}
     \end{center}
        \caption{The pressure $p$ at time $t=0$ in the computational domain is changed by the indentations in the tube walls. }
         \end{subfigure}
    \caption[The pressure $p$ at time $t=0$ and at $t=.1$s are shown in the computational domain. ]{The pressure $p$ at time (a) $t=0$  and at (b) $t=.1$s  are shown in the computational domain and its reflection over $r=0$.   The value of $p$ in the lower half of the domain is computed by imposing symmetry.}
    \label{fig:NScylP}
\end{figure}

The $w$ component of the velocity and pressure $p$ at times $t=0$s and  $t=0.1$s in the computational domain and its reflection over $r=0$ are shown in Figs. \ref{fig:NScylW} and \ref{fig:NScylP}, respectively. Note that this method can sharply capture the discontinuities in pressure and the non-smoothness in the velocity. At the time $t=0$ s, the velocity has a parabolic profile, and the pressure is linearly decreasing inside the vessel. At the time $t=0.1$s, the pressure increases at a faster rate where the tube narrows and decreases at a faster rate where the tube widens. The velocity decreases as the tube narrows and increases as the tube widens in the center. Near the distal end of the tube, the velocity decreases again.



\section{Discussion}
We have presented a numerical method for simulating viscous fluid flow through an open tube. The model is formulated as an immersed boundary problem, with the channel spanning from one end of the computational domain to the other. We apply the method to the Stokes equations and the Navier Stokes equations.
The Stokes equations are solved using a method that is an extension of the immersed interface method, which requires the immersed interface to be closed. This method gives second-order accurate values by incorporating known jumps for the solution and its derivatives into a finite difference method. 
The Navier-Stokes equations are solved using the velocity decomposition approach. That approach decompose the velocity into a “Stokes” part and a “regular” part. The first part is determined by the Stokes equations and the singular interfacial force. The regular part of the velocity is given by the Navier–Stokes equations with a body force resulting from the Stokes part. The regular velocity is obtained using a time-stepping method that combines the semi-Lagrangian method with the backward difference formula.
Numerical examples are presented to demonstrate that, for both the Stokes and Navier-Stokes models, the method converges with second-order spatial and temporal accuracy. 

The development of the present method is motivated by our interest in simulating biological problems with flows through biological tubes \cite{hu2019functional,layton2018renal,hu2020sex,hu2021computational,hu2021sex,layton2015modeling,layton2016predicted,layton2018sglt2,layton2016computational,layton2016solute,li2018functional,edwards2014effects,layton2017adaptive,sgouralis2017renal,sgouralis2015renal,chen2010effects,chen2011mathematical,fry2014impact,fry2015impacts} or microfluidic devices.


\bibliography{wileyNJD-AMA}%

\end{document}